\documentclass[12pt, a4paper]{article}

\usepackage{graphicx}
\usepackage[hyphens]{url}
\usepackage{hyperref}
\hypersetup{
    colorlinks=true,
    linkcolor=black,
    citecolor=black,
    filecolor=black,
    urlcolor=black,
}
\usepackage{xcolor}
\usepackage{mathtools}
\usepackage{booktabs}
\usepackage{microtype}
\usepackage{amssymb,amsmath,amsthm}
\theoremstyle{definition}
\usepackage[round]{natbib}

\usepackage{caption,setspace}
\usepackage[a4paper,left=2cm,right=2cm,top=2.5cm,bottom=2.5cm]{geometry}

\newcommand{\+}[1]{\ensuremath{\mathbf{#1}}}

\newcommand{\tpose}{\mathsf{T}}
\newcommand{\bs}[1]{\boldsymbol{#1}}

\newtheorem{definition}{Definition}

\title{Simulation Framework for Realistic Large-scale Individual-level Data Generation with an Application in the Health Domain}
\author{Santtu Tikka$^1$\thanks{corresponding author: santtu.tikka@jyu.fi} \and Jussi Hakanen$^2$ \and Mirka Saarela$^2$ \and Juha Karvanen$^1$}
\date{%
    $^1$Department of Mathematics and Statistics, University of Jyvaskyla, Finland\\%
    $^2$Faculty of Information Technology, University of Jyvaskyla, Finland\\[2ex]%
}

\begin{document}

\maketitle




\abstract{We propose a framework for realistic data generation and simulation of complex systems and demonstrate its capabilities in the health domain. The main use cases of the framework are predicting the development of risk factors and disease occurrence, evaluating the impact of interventions and policy decisions, and statistical method development. We present the fundamentals of the framework using rigorous mathematical definitions. 
The framework supports calibration to a real population as well as various manipulations and data collection processes. The freely available open-source implementation in R embraces efficient data structures, parallel computing and fast random number generation which ensure reproducibility and scalability. With the framework it is possible to run daily-level simulations for populations of millions of individuals for decades of simulated time. An example on the occurrence of stroke, type 2 diabetes and mortality illustrates the usage of the framework in the Finnish context. In the example, we demonstrate the data-collection functionality by studying the impact of non-participation on the estimated risk models and interventions related to controlling the additional salt intake.}\\


\maketitle
\newpage

\section{Introduction}\label{sec:intro}

Simulation is an important tool for decision making and scenario analysis, and healthcare is an example of a field where the use of simulation can provide substantial benefits. Simulated health data allow us to predict the population level development of risk factors, disease occurrence and case specific mortality under the given assumptions. Then, the predictions can be used in evaluating the effects of different policies and interventions in medical decision making in a prescriptive analytics approach. Simulated data are important also for statistical method development because the underlying true parameters are known and the obtained estimates can be easily compared with them. 

A simulation may start with real individual level data and simulate the future development with different assumptions. However, real health data are associated with legal and privacy concerns that complicate and sometimes even prevent their use in method development. Therefore, modeling the current state and using only synthetic data in simulation is often a viable option for a researcher who would like to evaluate the performance of a new method. These considerations lead to a conclusion that a general-purpose simulation framework should be modular and support different use cases. 

Below we outline the fundamental features and properties that an ideal simulator and the simulated population should support from the viewpoint of  statistical method development and decision making. These features are related to practical aspects of the simulated data, the technical aspects of the implementation and the types of phenomena we are most interested in.

\emph{Support for Alignment} In a typical use case, the goal is to match the simulated population to a real population of interest in the aspects that are viewed as the most important relative to the research questions at hand. The process which this is accomplished by is typically referred to as alignment. In order to make alignment possible, the simulated system can be calibrated by using real data according to chosen measures of similarity. The synthetic population under consideration should be sufficiently large to be considered representative. These principles are to be considered throughout the simulation period including the initial state of the population.

\emph{Support for Manipulations} The simulator should support external interventions. We approach this in three ways. An intervention can be a change in the system itself by either adding or removing individuals or events. Alternatively, an intervention can represent a policy decision that indirectly affects the individuals over a period of time. Finally, the intervention can directly target a specific individual, a subset of individuals or the entire population. An intervention of this third type is the idealized manipulation defined by the do-operator in causal inference literature \citep{Pearl:book2009}. For example, one may study the causal effects of a risk factor by forcing all members of the population to have the specific risk factor. Furthermore, interventions enable the study of counterfactual queries.

\emph{Support for Data Collection} The simulator should contain a specialized component for user defined data collection processes. This allows for the simulation of real world sampling procedures or selective participation where only a smaller subset of the entire population is available for practical analyses, e.g., via stratified sampling where different age groups have different sampling probabilities. Since the true simulated population is always available, the data collection feature can be used to assess the properties of data-analytical methods under a variety of conditions such as missing data that are prevalent in real-world data sets. Additionally, this feature can be used to assess the optimality of various study designs.

\emph{Support for Scalability} The simulated system can be scaled up in all aspects without incurring unreasonable computational burden. Large populations should be supported both in number of individuals and status variables through long simulation runs. Similarly, there should be no strict limitations in the number of possible events. Ideally, the system can be efficiently parallelized while minimizing the resulting overhead.

\emph{Support for Reproducibility} Any run of the simulator and sampling processes associated with it should be replicable. This allows for the replication of any statistical analysis carried out on data obtained from the simulator. The same requirement applies for simulation runs that take advantage of parallelized processing.

There exist open-source microsimulation frameworks developed in Python \citep{deMenten2014liam2}, 
Java \citep{richiardi2017jas,mannion2012jamsim,kosar2014simsalud,tomintz2017simsalud}, R \citep{richiardi2017jas,mannion2012jamsim,clements2021discrete}, and 
C++ \citep{clements2021discrete}. However, we found that no single simulation framework was able to encapsulate all of the previous features simultaneously or was otherwise unsuitable specifically for our purposes. For example, the \texttt{microsimulation} package in R by \citet{clements2021discrete} while open-source, does not include direct support for manipulations and a new simulator has to be constructed for each interventional scenario. A more comprehensive overview of the open-source frameworks that were included in our literature review is provided in Table~\ref{tab:relatedWorkOSframeworks} of Appendix~\ref{app:literature}.


In this paper, we propose an open-source simulation framework which we call \emph{Sima} to address the identified limitations of the existing frameworks. We present the fundamentals of the framework using rigorous mathematical definitions. In our approach, we adopt for simulation at the individual level. Each individual of the population is subject to a number of events that can for example describe changes in behaviour, occurrence of disease or illness, or mortality. The simulation proceeds using a fixed-increment time progression and the occurrence of events is controlled with stochastic processes. Thus, our simulation approach is an example of discrete event simulation. The efficient implementation enables high time resolution and makes it in some cases possible to approximate continuous-time processes. 

In addition to formulating the mathematical basis, we describe how the framework can be used in simulating complex systems, data collection procedures, and interventions in the health domain. An example on the occurrence of stroke, type 2 diabetes and mortality illustrates the usage of the framework in the Finnish context and studies the impact of non-participation on the estimated risk models. The simulation framework can be utilized in medical decision making where policies and interventions are optimized under multiple conflicting objectives. We see this as a topic for its own paper and concentrate here on describing the functionalities of the framework. 

The Sima framework has been implemented by using the R programming language. The source code of the implementation is available on GitHub at \url{https://github.com/santikka/Sima/} and extensive documentation can be found on the package website at \url{https://santikka.github.io/Sima/}.

The paper is organized as follows. \autoref{sec:framework} describes the simulator and its core components such as events and the population. Nuanced practical aspects such as calibration and interventions are also described. \autoref{sec:sampling} describes the data-collection process. \autoref{sec:implementation} presents the details of the implementation including the chosen data structures and libraries as well as optimization tools for calibration and other purposes. \autoref{sec:examples} illustrates a comprehensive application of the simulator for modeling the occurrence of stroke and diabetes with associated mortality. \autoref{sec:discussion} concludes the paper with a discussion and possible directions for future work.

\section{Simulation Framework}\label{sec:framework}

In this section we present the key concepts that define our simulation framework in detail. In order to accomplish the goals we have set out for our ideal simulator, we take into account the fundamental features of the previous section as explicitly as possible. Formal details of concepts presented in this section are available in Appendix~\ref{app:maths}. Figure~\ref{fig:flowchart} shows a flowchart that depicts the operation of a simulator constructed within the framework.

\begin{figure}[ht]
  \centering
  \includegraphics[width=0.92\textwidth]{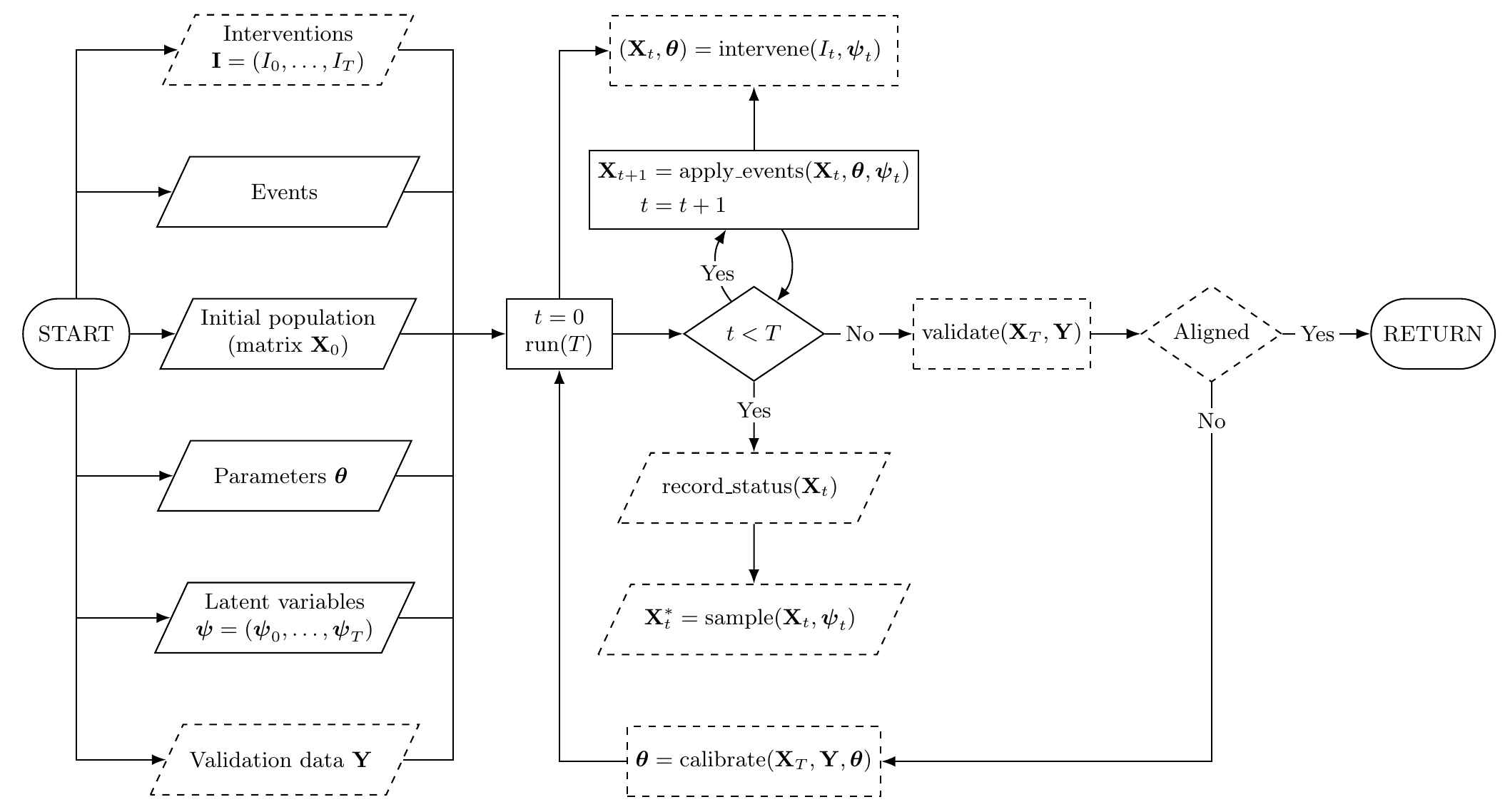}
  \caption{Flowchart of a simulator constructed within the Sima framework. Slanted nodes are inputs and outputs, rectangular nodes are processes and diamond nodes are decisions. Dashed nodes depict optional components. At the ``RETURN'' node, we return all generated outputs, meaning the final status of the population $\+ X_T$ and any samples $\+ X^*_t$ or entire records of the status $\+ X_t$ that may have been collected during the simulation. The node including ``calibrate'' should be understood as a procedure that provides a new values for the simulator parameters $\bs \theta$ based on the results of current run and the validation data.}
\label{fig:flowchart}
\end{figure}

The main operation of the simulator is built on four components: a population, a collection of events, a vector of parameters $\bs \theta$ and a vector of latent variables $\bs \psi$, each of which can change over time. A population is essentially a collection of status variables, which record various details of the simulated individuals, such as age, gender or disease events. Throughout the paper we assume that individual status variables are one dimensional and real-valued. This assumption does not rule out for example categorical variables, which can be included by encoding them in the standard way with a binary indicator for each category. Events govern the changes that may occur in the status of the population between time points. The parameters $\bs \theta$ define the functionality of the events, and the latent variables $\bs \psi$ are used to introduce random variation. For example, $\bs \theta$ might define the transition probabilities between states in a simple three-state model with the states ``healthy'', ``sick'' and ``dead'' and together the transition probabilities and a random value of $\bs \psi$ define the next state for each individual.

\subsection{Events}
 We draw a distinction between two types of events, ones that only modify the status variables of the existing individuals in the population and those that introduce new individuals into the population. We call these \emph{manipulation events} and \emph{accumulation events}, respectively. In both cases, the functionality of the events is based on the parameters $\bs \theta$ and the latent variables $\bs \psi$ specific to the simulator at hand.

Manipulation events are used to describe deterministic changes between two time points in the status variables of the population, either on the individual level or for a collection of individuals simultaneously. A manipulation event can be used for example to model stroke incidents based on the status of the current risk factors according to a probabilistic risk model. Our interest lies in the general effects of policy decisions over longer periods of time, and thus we do not focus on phenomena where individuals interact directly, such as the spread of viral infections in a network of individuals. However, there is no fundamental restriction to non-interacting populations, or manipulation events that only operate on the individual level. Interacting individuals can be hindrance on the scalability of the simulation, since a system of interacting individuals cannot be parallelized simply by partitioning the population appropriately unlike in the case of independent individuals.


Accumulation events are used to introduce new individuals into the simulated population. For example, an accumulation event could be used to describe incoming patients of a hospital. Initial members of the population and those added by accumulation events are never removed from the population. In the context of the hospital example, we still want to keep the records of discharged patients. However, a manipulation event may change the status of an individual in such a way that no further manipulation events have any affect on the status of the individual (such as mortality, or patient discharge, for example). However, a simulated sampling process may encode missing data in such a way, that information on all status variables of a specific individual or a set of individuals is not available in the resulting sample.

The events are defined as deterministic functions given the parameters and the latent variables which, if assumed to be random variables, give a stochastic nature to the simulator. Another important aspect is the order of events. Since manipulation events map the vector of status variables of a single individual into another vector of status variables, the order of application is significant. For accumulation events, the order of application simply corresponds to a permutation of rows in the resulting population, meaning that the order does not matter.

\subsection{Simulator}

To live up to its name, a simulator must have the capacity to advance the initial population through the application of manipulation events and accumulation events. At a given time~$t$, the simulator occupies a state $S_t$, which is an encapsulation of its current population and the parameter configuration. Transitions between states are carried out by the transition function which is defined by the events of the simulator. In turn, the functionality of the events is governed by the current state, the parameters $\bs \theta$ and latent variables $\bs \psi$. 

The transition function is the main workhorse of the entire simulation framework. Given a vector of latent variables, it defines the canonical transformation from one state to the next, i.e., from $S_t$ to $S_{t+1}$. This transformation consists of two parts, corresponding to the manipulation events and the accumulation events of the simulator. Since accumulation events do not depend on the current population, the order in which they are applied does not matter. However, the order is significant for manipulation events. Furthermore, this order can vary between individuals an can be controlled by the user. In a practical implementation, this can become a major performance concern if the population is large. Generating a random order for the manipulation events for every individual requires more computational effort than generating a single order for the entire population. 

As output of a single run of the simulator, the user is provided with a simulation sequence $\{S_t\}_{t=0}^T$ of desired length $T$. The simulation sequence begins from an initial state $S_0$, defined by the user, and a collection of latent variables obtained via pseudo-random number generation (PRNG). This sequence consists of all the states occupied during the simulation, allowing the study of the status variable processes. There are no restrictions for the time resolution. Note that in a single simulation sequence the parameters are determined by the state at the initial time and remain unchanged throughout the sequence.

\subsection{Alignment and Calibration}

To ensure realistic and sensible output, simulation output is typically aligned to some external targets \citep{Harding2007,LiODonoghue2014,Stephensen2015LogitSA}. Such targets may include joint or marginal distributions of specific status variables, expected total or accumulative event occurrences or event parameter values. Our primary approach to alignment is calibrating the simulator with respect to external data. For this purpose, we need to be able to change its parameter configuration while keeping the system otherwise intact. This feature is provided by the configuration function which manipulates the current state by altering the values of the parameter vector. More generally, calibration is based on optimization that involves multiple simulation runs, where the vector of latent variables is kept fixed between runs but the parameters are adjusted between each run according to some optimization strategy.

Calibration is an example of parameter estimation where the best parameter values of the model are found by using an optimization algorithm. Calibration of a simulator can be formulated as an optimization problem
\[
  \underset{\bs \theta \in \Theta}{\text{minimize }} g(\+ Y, \{\+ Z_t(\bs \theta)\}_{t=0}^T),
\]
where $g$ is the objective function (with possibly multiple objectives), $\+ Y$ represents external validation data and $\{\+ Z_t(\bs \theta)\}_{t=0}^T$ is a sequence of a set of status variables from a simulation sequence of length $T$ such that the initial state is defined by the user. The state of the latent variables $\bs \psi_t$ is kept fixed for different values of $\bs \theta$. 

The main idea is to make the model compatible with some external validation data. Thus, the objective function should be chosen such that it measures deviation in population level characteristics between the external validation data and the simulation output instead of properties of specific individuals. The quality of the calibration is thus dependent on a number of user-dependent factors. The chosen objective function, quality of external data, optimization method used and length of the simulation sequence should be carefully considered on a case-by-case basis. A common choice for the objective function is a least squares function where the deviation is measured as
\[
  g(\+ Y, \{\+ Z_t(\bs \theta)\}_{t=0}^T) = g(\+ Y, \widehat{\+ Y}(\bs \theta)) =
  ||\+ Y - \widehat{\+ Y}(\bs \theta)||^2,
\]
where $\widehat{\+ Y}(\bs \theta)$ is a function of the simulation output corresponding to the form of the validation data. In some cases, it may be necessary to scale up the simulation (e.g., increase the number of individuals or extend the length of the simulation) in order to obtain better calibration. Further, other types of deviation measures can be used instead of least squares depending on the properties of data used for calibration. In case of multiple calibration targets, the function $g$ can be multiobjective (see, e.g., \cite{enns2015identifying} for more details).

As an extension, we may also consider constrained optimization, where some properties of the simulation output are strictly enforced via the optimization procedure. This is closely related to alignment methods which aim to ensure that the output of the simulation conforms to some external criteria by adjusting the output itself, and not the model parameters. In our framework, this kind of alignment can be accomplished via suitably constructed manipulation events that are then applied either as a post hoc correction, or during the simulation. In practice, this is accomplished via interventions, which are discussed in the next section.

\subsection{Interventions and Counterfactuals}

Recalling our requirement for manipulability as outlined in \autoref{sec:intro}, the simulator is imbued with the capacity to allow for different types of interventions. The first type concerns a manipulation that either adds or removes individuals or events. This can be accomplished by constructing time-dependent events. A population change at a specific time can be either an accumulation event to add individuals, or a manipulation event to remove individuals by changing the value of status variable that indicates the presence of that individual in the population. Events on the other hand can be constructed in such a way that they only occur at specific time points during the simulation.

The second type of intervention is policy decisions. For example, consider a treatment regime consisting of two different treatments where the decision about the treatment to be prescribed is based on some threshold value of a health indicator, such as blood pressure. A policy decision could involve a change in this threshold value. Supposing that a parameter in $\bs \theta$ is the threshold value, a suitable intervention can be modeled by the configuration function such that only the specific parameter value is changed accordingly.

Finally, the last type of manipulation is the intervention encapsulated by the do-operator. In causal inference, the do-operator enforces a set of variables to take constant values instead of being determined by the values of their parents in the associated causal graph \citep{Pearl:book2009}. When applied to a simulated population, an intervention of this type enables the direct study of causal relationships. For example, we can study the effect of how changing precisely one feature in the entire population affects the outcomes of interest. To accurately simulate the do-operator, the events corresponding to mechanisms that are desired to be subject to intervention have to be defined in a way that encodes the intervention as a parameter in $\bs \theta$. This allows the use of the configuration function to enable the desired actions on the events themselves.

The do-operator also allows for the direct study of counterfactual queries. Consider again the treatment regime with two different treatments. In the simplest case, these two potential outcomes correspond to treatment and control, where no treatment is prescribed. In real world scenarios, it is impossible to observe both potential outcomes simultaneously from the same study unit, which is known as the fundamental problem of causal inference. We can bypass this problem with ease by simulating two populations that are identical in all aspects with the exception of the treatment assignment while using the same vector of latent variables. More generally, we can apply this approach to study any number of potential outcomes and parallel worlds \citep{Pearl:book2009}.

\section{Data Collection Process} \label{sec:sampling}

Simulating data collection adds another dimension to the simulation framework. Conceptually, the idea is the same as in causal models with design where the causal model (event simulator) and the data collection process together define the observed data \citep{Karvanen2015studydesign}. Data collection is important especially in method development because statistical methods aim to estimate underlying processes on the basis of limited data. Usually, data on most health variables can not be collected daily. For instance, it is not realistic to assume that measurements requiring laboratory testing of a blood sample would be available for a population cohort on every day. In a typical longitudinal study, most of daily measurements are in fact missing by design. Using data from a small number of time points to learn about the underlying continuous-time causal mechanism may lead to unexpected results if the data collection process is not appropriately accounted for \citep{aalen2016can,strohmaier2015dynamic,albert2019continuous}.

Even if the data are planned to be collected, things may go wrong. Real data are almost never of perfect quality. Data may be incorrectly logged, completely lost or otherwise degraded. Some values or entire study units may suffer from missing data or measurement error. In surveys and longitudinal studies non-participation, dropout and item non-response are always possible; a problem which cannot always be remedied even by recontacting the participants that dropped out or decided not to answer the survey. More generally, participation in a study may be influenced by a number of factors resulting in selective participation. The aforementioned aspects are also an important component of our simulation framework. Obtaining a smaller subset than the full population after or during the simulation is an important feature for many data-analytical tasks. The presented framework includes a highly general interface for simulating a realistic sampling process from the synthetic population.

In the presented framework, we are able to construct missing data mechanisms of any type, especially those recognizable as belonging to the traditional categories of missing completely at random, missing at random and missing not at random \citep{littlerubin2002}. Having constructed the mechanism, we may obtain a sample from the full population where the mechanisms dictate the values or observations that will be missing in the sample. Multiple imputation is a widely applied and studied method for replacing missing data with substituted values \citep{vanbuuren2018flexible}. The simulated sample with missing data can be used to assess the possible bias introduced by imputation and the representativeness of any results obtained from the imputed data by comparing the results to those obtained from the same sample without any missing data.

\section{Implementation} \label{sec:implementation}

We chose R as the programming language and environment to be used for the implementation of the simulator \citep{Rsoftware}. R provides a vast multitude of computational and statistical methods as open-source packages via the Comprehensive R Archive Network (CRAN). Currently, the CRAN repository features over 15\,000 packages that encompass methodology from the full variety of scientific disciplines including physics, chemistry, econometrics, ecology, genetics, hydrology, machine learning, optimization and social sciences. This is especially useful when constructing manipulation events, since statistical models and other methods from state-of-the-art research can be directly implemented via events in the simulator to accurately model phenomena of interest. After carrying out the simulation, the synthetic data can be effortlessly exported from R to various external data formats to be used in other applications, if desired.

\subsection{Data Tables}

We opted for data tables implemented in the \texttt{data.table} package to act as the container for the population \citep{datatable}. Compared to standard R data containers, data tables have a number of advantages such as efficient subset construction, in-memory operations, internal parallelization and a syntax that resembles that of SQL queries. Data tables are also highly memory-efficient and capable of processing data with over one billion rows. The choice was further motivated by the extremely competitive performance of data tables compared to other popular open-source database-like tools in data science. A regularly updated operational benchmark can be found at \url{https://h2oai.github.io/db-benchmark/} that measures the speed of data tables and other popular packages such as the \texttt{datatable} Python package and the \texttt{DataFrames.jl} Julia package in carrying out various tasks with input tables up to a size of one billion rows and nine columns (translating to 50GB in memory). Internally, the \texttt{data.table} package does not depend on any other R packages, making it a future-proof choice and easily extendable by the user.

Operational performance of data tables is exemplified both in carrying out computations within groups in a single table as well as joining multiple tables. For our purposes, efficient operations within a single table is of primary interest. As an example, consider a manipulation event that specifies mortality due to stroke as a logistic regression model. In a data table, the event probabilities for the group that has had at least one stroke can be efficiently computed using vectorized covariates (a subset of the status variables) without making any unnecessary copies of the objects involved.

\subsection{Reference Classes}

Manipulation events, accumulation events and the simulator itself are defined as R6 reference classes. R6 is an object oriented programming system for R provided through the \texttt{R6} package \citep{Rsix}. R6 does not have any package dependencies and does not rely on the S4 or S3 class systems available in base R unlike the default reference class system. R6 features several advantages over the built-in reference class system such as vastly improved speed, ease of use and support for public and private class members. The primary reason to adopt a reference class system is the mutability of the resulting objects; they are modified in place unlike standard R objects, which are always copied upon modification or when used in function calls. See the \texttt{R6} package vignette titled ``Performance'' for speed and memory footprint comparisons between R6 and base R reference classes (also available at \url{https://r6.r-lib.org/articles/Performance.html}).

The typical structure of an R6 class object consists of both fields and methods. Following the standard object-oriented programming paradigm, fields and methods can be either public or private depending on whether they should be accessible externally or only within the class object itself. It is also possible to construct so called active fields that are accessed as fields but are defined as functions, just like methods are.

Classes for both manipulation events and accumulation events are inherited from a superclass that is simply called \texttt{Event}. An \texttt{Event} can be given a name and a short description to provide larger context. More importantly, an \texttt{Event} contains the fields \texttt{mechanism} and \texttt{par}, which determine the functionality and parameter configuration of the \texttt{Event}, respectively. The \texttt{mechanism} can be any R expression and its functionality may depend on the parameters defined by \texttt{par}. The \texttt{Event} superclass provides a template method called \texttt{apply} to evaluate the \texttt{mechanism}, but the actual behaviour of this method is determined by the type of the event and fully controlled by the simulator. The inherited class \texttt{ManipulationEvent} defines the type of events of its namesake. For events of this class, the \texttt{apply} method provides access to the current status of the simulated population and evaluates the \texttt{mechanism} in this context, but it is not allowed to introduce new individuals, only to modify the existing ones. On the other hand, the inherited class \texttt{AccumulationEvent} allows its \texttt{mechanism} to add new members to the simulated population via the \texttt{apply} method, but is prohibited from manipulating the current population.

The \texttt{Simulator} class implements all of the primary functionality of the framework including population initialization, state transitions, calibration and interventions. The constructor of the class takes lists of \texttt{ManipulationEvent} and \texttt{AccumulationEvent} objects as input and calls a user defined function \texttt{initializer}, which constructs the initial population. The \texttt{run} method is the transition function of the simulator and it uses the \texttt{apply} methods of the event objects given in the constructor at each time point to advance the simulation. The user may control how many state transitions are carried out sequentially for a single \texttt{run} command. The simulator can be instructed to conduct the state transitions in parallel by calling the \texttt{start\_cluster} method beforehand. This method connects the simulator to a user defined parallel backend such as an MPI cluster initialized using the \texttt{doMPI} package and distributes the population into chunks. Parallel computation can be stopped with the \texttt{stop\_cluster} method which also merges the individual population chunks back together. Configuration of the event parameters is managed by the \texttt{configure} method which takes as input the amount of state transitions to be used, the set of parameters to be configured and an output function to be used as a part of the objective function in the optimization for calibration. The output function can be simply a statistic that is obtained from the resulting simulation sequence to be compared to a set of validation data. For an example, see the objective function defined for calibration in Section~\ref{sec:example_calibration}. Interventions can be applied at any point of the simulation via the \texttt{intervene} method. This method can apply a set of manipulation events, accumulation events, or parameter configurations to change the current state of the simulator.

\subsection{Parallel computation}
The task of simulating the synthetic population in the described framework can be considered ``embarrassingly parallel'' \citep{herlihy2011art}. Communication between the parallel tasks is required only when there is a need to consider the population as a whole, such as for calibration or for constructing a time series. In the ideal case, the end state of the population is solely of interest which minimizes the required communication between tasks.

Parallel CPU processing functionality is provided by the \texttt{doParallel} and \texttt{foreach} packages \citep{doParallel,foreach}. These packages allow for parallel computation on a single machine with multiple processors or cores, or on multiple machines. Alternatively, Message Passing Interface (MPI) can be used through the packages \texttt{doMPI} and \texttt{Rmpi} \citep{doMPI, Rmpi}. More generally, any parallel adaptor to the foreach looping construct is supported. The parallel approach takes full advantage of the independence assumption of individuals. This allows us to partition the population into chunks, typically of equal size at the start of the simulation run, and to evaluate the events in each chunk separately by distributing the chunks to a desired number of workers. Shared memory between the workers is not required, and the full unpartitioned population does not have to be kept in memory during the simulation even for the supervisor process. The \texttt{dqrng} package allows for multiple simultaneous random number streams, which in turn enables the use of different streams for each worker and makes the parallel simulation sensible. The initial seeds for these streams can be manually set by the user for reproducibility. For more details on PRNG in the implementation, see Appendix~\ref{app:dqngr}.

Additional overhead in both memory consumption and processing time is incurred by having to distribute metadata on the population and other necessary information to the subprocesses such as variable names and the \texttt{Event} instances. Furthermore, constructing the chunks and reintegrating them back together at the end of simulation is an added hindrance. However, these operations do not depend on the length of the simulation run meaning that the overhead from these particular tasks becomes negligible as the length of the run grows. 

\begin{figure}[ht]
  \centering
  \includegraphics[width=0.480\textwidth]{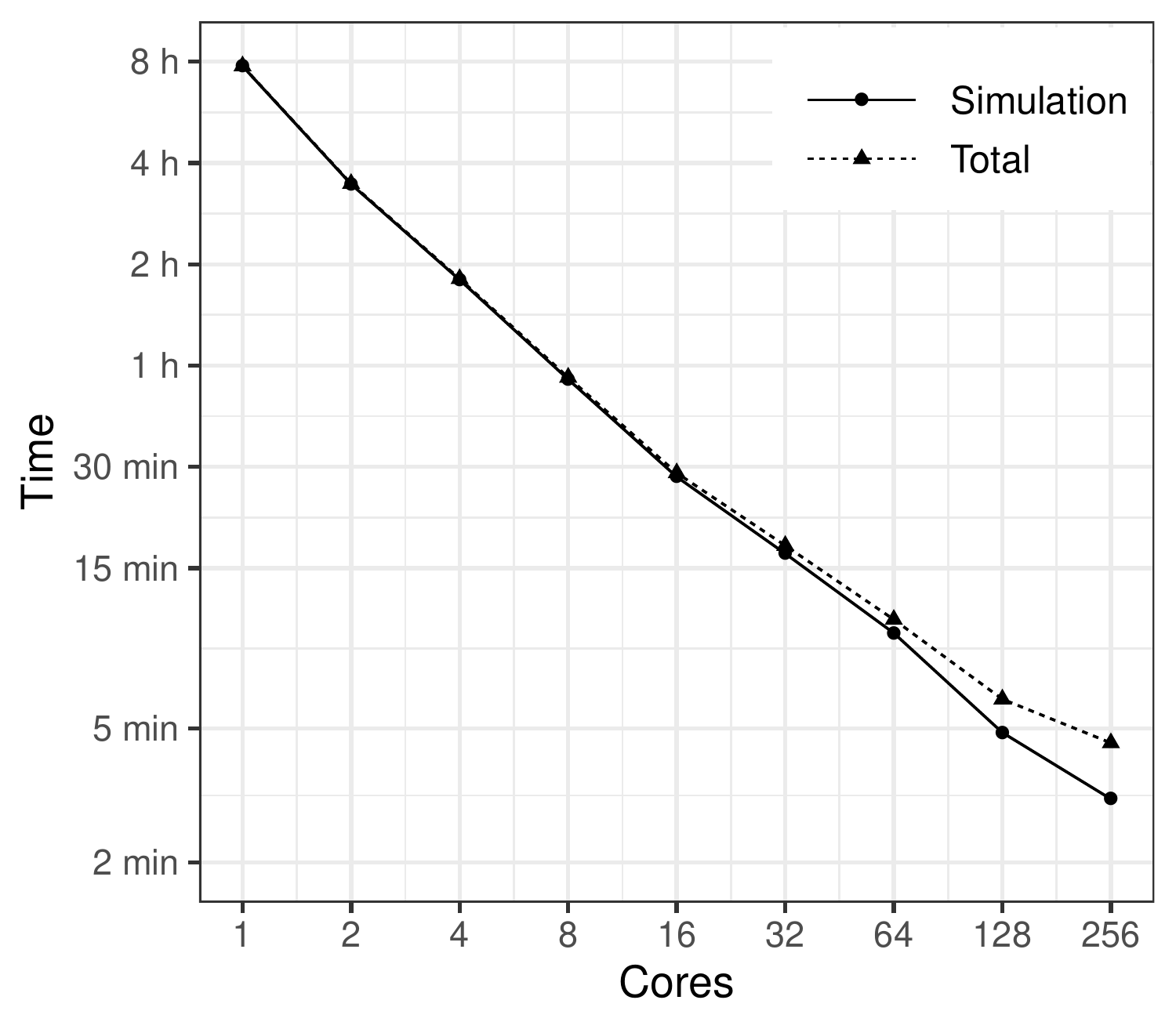}
  \caption{Performance benchmark using a population of 10 million individuals, 5 events and a simulation period of 10 years on a time scale of one day. Total time also includes the time it takes to generate the initial population, initialize the parallel computation cluster and to distribute the data to the cores. MPI was used to conduct the benchmark. }
\label{fig:scalability}
\end{figure}

Figure~\ref{fig:scalability} shows how the processing time is affected when the number of computer cores is increased. The benchmark scenario consists of a population of 10 million individuals, 5 events and a simulation period of 10 years where the events are applied daily. More specifically, the scenario is the one described in Section~\ref{sec:examples}. The results show that the runtime of the simulation is cut almost in half every time the number of cores is doubled (geometric mean of the consecutive runtime ratios is 0.534). As expected, the proportion of the overhead resulting from the communication between cores and other parallel tasks also grows as the number of cores is increased. It should be noted that in the runs involving up to 32 cores only one computing node was used. For the runs with 64, 128 and 256 cores, the tasks were distributed to three, four, and ten computing nodes, respectively.

\subsection{Optimization}
In this paper, optimization is used for calibration of the simulator parameters $\bs \theta$. The optimization method used needs to be selected based on the properties of the objective function used for calibration. There exists many optimization tools implemented in R and we briefly describe the \texttt{optim} function that provides a variety of optimization algorithms. The \texttt{optim} function is developed for general purpose optimization where box constraints can be set for the decision variables. The function includes implementations of the Nelder-Mead, quasi-Newton, and conjugate gradient algorithms. In the example of Section~\ref{sec:examples}, we have used the least squares function which is differentiable, and in the corresponding calibration problem, we do not have explicit bounds for the calibrated parameters and we don't calculate any gradients for the objective function. Therefore, we will use the Nelder-Mead algorithm from the \texttt{optim} function which is suitable for our purposes \citep{NelderMead1965}. The Nelder-Mead algorithm doesn't use any gradient information but works on a set of $N+1$ points in a optimization problem having $N$ decision variables. The main idea is at each iteration to improve the objective function value of the worst point by using reflections and contractions to identify locally optimal points.

\section{Example on Diabetes and Stroke} \label{sec:examples}

We present an example use case of the simulator where we model the occurrence of stroke, type 2 diabetes and mortality in a synthetic population of 3.6 million individuals, corresponding to the Finnish population size of persons over 30 years of age in the year 2017. The goals of the illustration are threefold. First, we wish to exemplify the use of the simulation framework  in building simulators. Second, we aim to build a simulator that can realistically model the occurrence of stroke, type 2 diabetes and mortality in the Finnish population. Third, we want to illustrate to use of the simulator in methodological comparisons and carry out a simulation where we study the impact of non-participation on the estimated risk model of stroke. See \url{https://santikka.github.io/Sima/articles/example.html} for the annotated R codes and data files used in the example.

We generate a synthetic population based on the population of Finland that includes the status of following risk factors for each individual: smoking, total cholesterol,  high-density lipoprotein (HDL) cholesterol, systolic blood pressure, age, sex, body mass index, high blood glucose, waist circumference, diabetes and whether either parent has suffered a stroke. A single state transition in the simulator corresponds to one day in real time. The events include occurrence of stroke, occurrence of diabetes, death after stroke and death without suffering a stroke.  The events modify status variables that describe the health status of the individual. The daily probabilities of the events depend on the risk factors. The goal is to create a synthetic follow-up study about this population with selective non-participation and use the obtained sample to estimate the effect of the risk factors on the risk of stroke and diabetes. As the daily event probabilities are low, it is unlikely that a person has more than one event per day and the simulation can be therefore used to study competing risks.

\subsection{The Models for the Events} \label{sec:examplemodels}

Mortality is modeled separately for individuals that have suffered a stroke and for those that have not. A stroke is considered fatal if an individual dies within 28 days of the initial stroke incident. For simplicity, it is assumed that there can be at most one stroke incident for any individual. The long term stroke survival rates are modeled according to the results by \citet{stroke_survival1}. and the remaining mortality is calibrated to match the Finnish population in the year 2017 based on two factors: mortality rates in each age group and the proportion of deaths due to stroke out of all deaths over one year. We use the official mortality statistics from Statistics Finland for this purpose \citep{mortality}.

The model for the occurrence of stroke is based on the FINRISK 10-year risk score \citep{VARTIAINEN2016213}. In the risk score calculation, the 10-year risk of cardiovascular disease is modeled as a sum of two risks: the risk of coronary heart disease and the risk of stroke, both of which are modeled via logistic regression. For our purposes, we use only the stroke portion of the risk score model. The 1-day risk for the corresponding manipulation event in obtained by assuming a Poisson process for the stroke. Similarly, the model for the occurrence of type 2 diabetes is based on the FINDRISC (Finnish Diabetes Risk Score) 10-year risk score \citep{FINDRISK}. Both the FINRISK and the FINDRISC risk score models take into account a number of risk factors. Since these models only consider individuals over the age of 30, we also restrict our simulated population in the same way in terms of age.

The 10-year risk of stroke is modeled separately for both genders while including the same risk factors. More specifically, the risk factors for stroke include age, smoking status, HDL cholesterol, systolic blood pressure, diabetes status and an indicator whether either parent of the individual has suffered a stroke. We let $C_{it}(t^*)$ denote a Poisson process counting the number of stroke events that have occurred after $t^*$ days since time $t$ for each individual $i$ giving us the following relation
\[
       \textrm{logit}(\mathbb{P}(C_{it}(3650) > 0)) = 
                     \mathbb{I}(G_i = 0) \+ x^\tpose_{it} \bs \beta_m + \mathbb{I}(G_i = 1) \+ x^\tpose_{it} \bs \beta_f,
\]
where $G_i$ denotes the gender (1 = woman), $\+ x^\tpose_{it}$ is the row vector of status variables for individual $i$ at time $t$, and $\mathbb{I}$ is an indicator function. The parameter vectors for the stroke model are denoted by $\bs\beta_{m}$ and $\bs\beta_{f}$ for men and women, respectively. The values of the model parameters are obtained from the FINRISK risk score. From the 10-year risk we first obtain the intensity $\tau_{it}$ of the underlying Poisson process via the following relation
\[
\mathbb{P}(C_{it}(t^*) = m) = \frac{(\tau_{it} t^*)^m}{m!} e^{-\tau_{it} t^*}.
\]
When applied to the 10-year risk we have
\[
\tau_{it} = -\frac{\log(\mathbb{P}(C_{it}(3650) = 0))}{3650}.
\]
The one day risk of stroke is now modeled as
\[
  \textrm{logit}(\mathbb{P}(S_{it} = 1|S_{it} = 0)) = 1 - \exp(-\tau_{it}),
\]
where $S_{it}$ is an indicator denoting whether individual $i$ has suffered a stroke at time $t$ or earlier.

The one day risk of diabetes is obtained in the exact same way from the model of the 10-year risk of diabetes as the risk of stroke. For the diabetes FINDRISC model, the set of risk factors is different, and the same model parameters are now used for both genders. The risk factors used in the model are age, body mass index, waist circumference, use of blood pressure medication and high blood glucose.

We use the findings by \citet{stroke_survival1} to model the long-term survival after stroke. The study was conducted as a part of the World Health Organization (WHO) MONICA (Monitoring Trends and Determinants in Cardiovascular Disease) project during 1982--1991 and consisting of subjects aged 25 years or older in the Glostrup region of Copenhagen County in Denmark. In the study, a number of stroke subtypes were considered, but for our purposes we simply use the aggregate results over all variants to model the mortality due to stroke.

The cumulative risk of death reported by Br{\o}nnum-Hansen et al. is 28\%, 41\%, 60\%, 76\% and 86\% at 28 days, 1 year, 5 years, 10 years and 15 years after the stroke, respectively. Since the survival probability decreases rapidly during the first 28 days, we construct two models taking into account this 28 day period, and the remaining period up to 15 years from the first stroke incident. The probability that an individual $i$ who has suffered a stroke does not die at time $t$ given that they are still alive at time $t-1$ and have survived for over 27 days is modeled using a regression model with a logistic link
\begin{equation} \label{eq:stroke_survival}
    \textrm{logit}(\mathbb{P}(Y_{it} = 0|Y_{i,t-1} = 0, S_{it} = 1)) = 
    \alpha_1 + \alpha_2 \exp(\alpha_3 (t_i - 27)), \quad t_i > 27,
\end{equation}
where $Y_{it} = 0$ denotes survival of individual $i$ at time $t$. The time from the stroke occurrence in days is $t_i$ for individual $i$. Values of the parameters $\alpha_1, \alpha_2$ and $\alpha_3$ are found by directly fitting the model of \eqref{eq:stroke_survival} to the reported survival rates.

The survival during the first 28 days from the stroke incident is assumed to be constant for each day
\[
   \textrm{logit}(\mathbb{P}(Y_{it} = 0|Y_{i,t-1} = 0, S_{it} = 1)) = 
   \alpha_0, \quad t_i < 28.
\]
The reported survival during the initial period varies drastically between the different stroke variants and evens off as more time passes from the first stroke incident. As our focus lies in simulating a period of 10 years, we consider this simple model to be sufficient for such a small time window of the entire simulation period.

The secondary mortality from other causes is modeled using a Weibull distribution with the following parametrization for the density function
\[
  f(x) = \frac{\lambda}{\kappa} \left(\frac{x}{\kappa}\right)^{\lambda-1}\exp\left(-\left(\frac{x}{\kappa}\right)^\lambda\right), \quad x > 0,
\]
where $\lambda > 0$ is the shape parameter and $\kappa > 0$ is the scale parameter of the distribution. The probability that individual $i$ dies at a specific time $t$ without having previously suffered a stroke given that they are still alive at time $t-1$ is defined by
\[
  \mathbb{P}(Y_{it} = 1|Y_{i,t-1} = 0, S_{it} = 0)
  = F(A_{it};\lambda,\kappa) \\
  = 1 - \exp\left(-\left(\frac{A_{it}}{\kappa}\right)^\lambda\right),
\]
where $A_{it}$ is the age of individual $i$ at time $t$. $F$ is the cumulative distribution function of the Weibull distribution. 

\subsection{Initial Population}

Equally important with the accurate modelling of the events is the generation of realistic initial values for the various health indicators, risk factors and other variables in the population. Each individual is defined to be alive at the initial state and the gender is generated simply by a fair coin flip. The ages are generated separately for both genders based on the age structure of Finland in the year 2017 from the official statistics with the restriction that every individual is at least 30 years old \citep{popstruct}.

We use Finnish data (the 20\% sample available as open data) from the final survey of the MONICA project to obtain initial values for BMI, waist circumference, cholesterol, HDL cholesterol and systolic blood pressure \citep{monographtunstall}. First, we transform these variables in the data into normality using a Box-Cox transformation separately in each age group, for both genders and based on smoking status \citep{box1964analysis}. Smoking is defined here as a dichotomous variable so that an individual is considered a smoker if their self-reported smoking frequency for cigarettes, cigars or pipe was either ``often'' or ``occasionally'' in the MONICA data. Next, we fit a multivariate normal distribution to the transformed data in each subgroup and generate values from this distribution in order to take the correlation structure of the variables into account. Finally, the inverse transformation is applied to the generated values. We also apply a post-hoc correction to the generated values by restricting them to the ranges observed in the Health 2000--2011 study data. Values that fall outside the valid ranges are replaced by other generated values that fell within the valid ranges from individuals in the same age group that are of the same gender and have the same smoking status.

The report by \citet{koponen2018} provides distributions of smoking and stroke in a representative sample of the general Finnish population in the year 2017. For example, the study reports the proportion of non-smokers for each age group starting from 30-year-olds with 10 year intervals by gender. For the FINRISK model, the variable that either parent of an individual has suffered a stroke is initialized by setting the probability of having at least one parent who has suffered a stroke to be the sum of the gender specific age adjusted stroke incidence rates for persons over 50 years old in the FinHealth 2017 study. For smoking, we use the reported gender and age specific proportions directly as parameters for binomial distributions to generate the initial values. For those individuals in the initial population that have suffered a stroke, an accumulated survival time is generated from a Weibull distribution fitted to the 1 year, 5 year, 10 year and 15 year survival rates that were reported in \citep{stroke_survival1} while ensuring that the stroke could not have occurred before 30 years of age.

After all other initial values have been generated, the initial diabetes status is generated by first using the FINDRISC model to predict the 10-year risk for each individual in the initial population, and then adjusting the individual risks such that the total expected gender specific diabetes prevalence matches that of Finland in the year 2017 (15\% for men and 10\% for women). The adjusted risks are then used as the probabilities of having diabetes at initialization for each individual.

\subsection{Calibration} \label{sec:example_calibration}

We employ calibration to obtain values for the parameters $\lambda$ and $\kappa$ of the non-stroke related mortality model, as well as for the parameter $\alpha_0$ of the fatal stroke model. The objective function $g$ compares the official mortality statistics to those obtained from the simulation for a period of one year. 
\[
  g(\+ Y, \widehat{\+ Y}(\bs \theta)) = \sum_{k = 30}^{99} w_k \left[\log(\hat{y}_k) - \log(y_k)\right]^2
                                      + w^* \left[\log(\hat{p}) - \log(p)\right]^2
\]
where $\widehat{\+ Y}(\bs \theta) = (\hat y_{30}, \ldots, \hat y_{99},\hat p)$ is the simulation output with the event parameters set to $\bs \theta$, $w_k$ and $w^*$ are weights, $\hat y_{30}, \ldots, \hat y_{99}$ denotes the estimated mortality for each age group $k$ and $\hat p$ is the estimated proportion of stroke deaths. $\+ Y$ denotes the corresponding true numbers $(y_{30}, \ldots, y_{99}, p)$ obtained from the official statistics. The weights $w_k$ were chosen so that $w_k = 1$ when $k < 80$ and $w_k = 100$ otherwise. In this case, we observed that the optimum is reached faster when using a weighted least squares approach, since the mortality is much higher in the older age groups. On the other hand, the total yearly mortalities for younger age groups are close to zero, meaning that the size of the simulated population has to be sufficiently large in order for the estimates to be accurate. The value $2000$ was chosen for the weight $w^*$ which roughly translates to equal importance of the age group specific mortalities and the proportion of strokes in the optimization.

\begin{figure}[ht]
  \centering
  \includegraphics[width=0.480\textwidth]{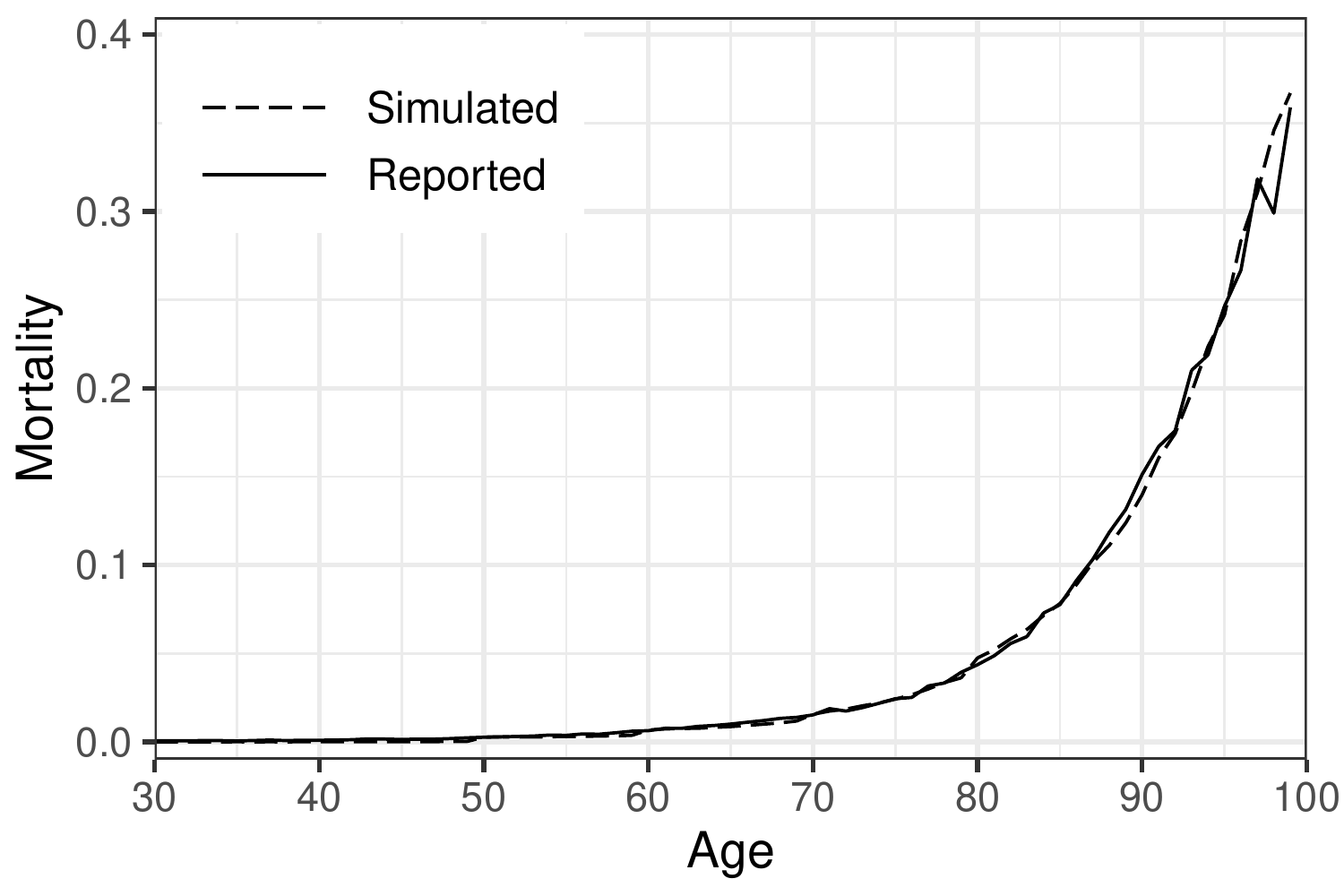}
  \caption{Simulated mortality in a population of 3.6 million individuals after calibration and official mortality statistics of Finland for the year 2017 by age. A population of one million individuals was used for the calibration. }
  \label{fig:mortality}
\end{figure}

We found that increasing the population size beyond one million individuals did not change the result of the calibration in this scenario. We also considered more elaborate weighting schemes, but the effect of the weights diminishes as the population size grows. The described weights were chosen based on trial and error with small populations (less than 10\,000 individuals). Figure~\ref{fig:mortality} shows the simulated mortalities after calibration for each age group as well as those obtained from the official statistics for ages 30--99. In the year 2017, 7.5\% out of all deaths in Finland were caused by a stroke (classification falling under ICD-10 Chapter IX I60--I69). For the synthetic population, the same number is 7.8\%.

\subsection{Sampling and Non-participation}

We simulate a data collection process where individuals from the simulated population are invited into a health examination survey. Measurements of the relevant risk factors are taken at baseline, and the study participants are then subsequently followed for the 10 year period. In real world studies, participation may depend on a number of factors especially when dealing with sensitive information about an individual's health. This leads to selective non-participation. We use a logistic regression model to simulate the effect that the relevant status variables have on the probability of non-participation at baseline. The chosen model is
\begin{equation} \label{eq:missigness}
    \mathrm{logit}(\mathbb{P}(M_i = 1)) =  \rho_0 + \rho_1 G_i + \rho_2 A_i + \rho_3 B_i + \rho_4 O_i + \rho_5 W_i
\end{equation}
where $M_{i}$ is an indicator of non-participation and $G_i$, $A_i$, $B_i$, $O_i$ and $W_i$ are the gender (1 = woman), age (in years), body mass index (BMI), smoking indicator and waist circumference (in cm) of individual $i$ at baseline, respectively. Estimates for the parameters $\rho_k$, $k=0,\ldots,5$ are based on data from the Health 2000--2011 study \citep{harkanen2016systematic} and are shown in Table~\ref{tab:odds_ratios} along with their 95\% confidence intervals.

\begin{table}[!ht]
    \centering
    \footnotesize
    \caption{Estimated parameter values, odds ratios and their 95\% confidence intervals for the model of baseline non-participation.}
    \begin{tabular}{lll}
        \toprule
                    & \multicolumn{1}{l}{\hphantom{$-$}$\widehat{\rho_k}$} & \multicolumn{1}{l}{$\widehat{\exp(\rho_k)}$} \\
        \midrule
        Intercept   & $-$9.48 ($-$10.68, $-$8.32)    &       \\
        Gender      & $-$0.31 ($-$0.62, $-$0.01)     & 0.73 (0.54, 0.99)  \\
        Age         & \phantom{$-$}0.11 (0.10, 0.12) & 1.12 (1.11, 1.13)  \\
        BMI         & $-$0.09 ($-$0.15,$-$0.03)      & 0.91 (0.86, 0.97)  \\
        Smoking     & \phantom{$-$}0.69 (0.44, 0.94) & 1.99 (1.56, 2.56)  \\
        Waist Circumference & \phantom{$-$}0.04 (0.02, 0.06) & 1.04 (1.02, 1.06) \\
        \bottomrule
    \end{tabular}
    \label{tab:odds_ratios}
\end{table}

\subsection{Simulation results on stroke incidence and risk factors}
After generating the initial population, we then proceeded to run the simulation for a 10 year period. From the initial population of 3.6 million individuals, a subset of 10\,000 individuals were randomly sampled to be invited into a health examination survey with non-participation occurring at baseline. Individuals who had already suffered a stroke were excluded.

From the initial invitees, 336 individuals were excluded due to having a previous stroke incident. Afterwards, 1\,644 invitees decided not to participate at all. In this scenario, we assume that there is no item non-response for the 7\,900 remaining participants. Similarly to the FINRISK risk score, we model the 10-year risk of stroke separately for both genders via logistic regression. We compare the results from A) the sample with baseline non-response, B) the sample without non-participation and C) the entire synthetic population.  Tables~\ref{tab:odds_ratios_risk_male} and \ref{tab:odds_ratios_risk_female} show the results for the parameter estimation for men and women, respectively. 


\begin{table}[!ht]
    \centering
    \footnotesize
    \caption{Estimated odds ratios and their 95\% confidence intervals for the model of 10-year risk of stroke for men.}
    \begin{tabular}{lllll}
        \toprule
                & A & B & C & FINRISK \\
        \midrule
        Age         & 1.11 (1.10, 1.13)          & 1.10 (1.10, 1.11)         & 1.10 (1.09, 1.10)         & 1.12 \\
        Smoking     & 2.13 (1.52, 2.96)          & 1.67 (1.27, 2.19)         & 1.53 (1.50, 1.55)         & 1.65 \\
        Systolic blood pressure          & 1.01 (1.00, 1.02)          & 1.01 (1.01, 1.02)         & 1.02 (1.02, 1.02)         & 1.02 \\
        HDL cholesterol         & 0.73 (0.48, 1.07)          & 0.67 (0.49, 0.91)         & 0.65 (0.64, 0.66)         & 0.64 \\
        Diabetes        & 2.30 (1.67, 3.13)          & 2.38 (1.87, 3.02)         & 2.31 (2.28, 2.34)         & 2.41 \\
        Parents' stroke         & 1.25 (0.83, 1.84)          & 1.41 (1.04, 1.89)         & 1.33 (1.31, 1.35)         & 1.34 \\
        \bottomrule
    \end{tabular}
    \label{tab:odds_ratios_risk_male}
\end{table}

\begin{table}[!ht]
    \centering
    \footnotesize
    \caption{Estimated odds ratios and their 95\% confidence intervals for the model of 10-year risk of stroke for women.}
    \begin{tabular}{lllll}
        \toprule
                & A & B & C & FINRISK \\
        \midrule
        Age         & 1.06 (1.05, 1.08)         & 1.05 (1.04, 1.06)         & 1.05 (1.05, 1.05)         & 1.07 \\
        Smoking        & 1.67 (1.01, 2.66)         & 1.62 (1.05, 2.44)         & 1.40 (1.37, 1.44)         & 1.52 \\
        Systolic blood pressure          & 1.01 (1.00, 1.01)         & 1.01 (1.01, 1.02)         & 1.01 (1.01, 1.01)         & 1.01 \\
        HDL cholesterol         & 0.50 (0.32, 0.76)         & 0.54 (0.37, 0.78)         & 0.45 (0.44, 0.46)         & 0.47 \\
        Diabetes        & 3.70 (2.54, 5.34)         & 3.37 (2.43, 4.62)         & 3.21 (3.16, 3.27)         & 3.45 \\
        Parents' stroke         & 2.18 (1.42, 3.27)         & 2.02 (1.39, 2.87)         & 1.69 (1.65, 1.72)         & 1.73 \\
        \bottomrule
    \end{tabular}
    \label{tab:odds_ratios_risk_female}
\end{table}

The incidence of stroke in sample A was 543 of 100\,000 per year whereas in sample B the same number is 728 of 100\,000 per year, showing the bias caused by the baseline non-response, since the incidence in the full population in sample C is 723 of 100\,000 per year. As expected, the odds ratio estimates do not differ greatly between samples A and B, since the missingness mechanism of \eqref{eq:missigness} is ignorable as it does not depend on the stroke occurrence itself.

While some discrepancies remain between the true values and the estimates even when using the entire synthetic population, namely in the case of smoking and diabetes, overall the results indicate that adapting the one day risk from the 10-year risk by assuming a Poisson process is a valid strategy in this instance. The example further highlights the importance of taking the underlying correlation structure of the risk factors into account in the initial data generation. For example, the MONICA data does not contain information on diabetes, so some loss of information likely occurs when the initial diabetes status is based purely on the FINDRISC model. Similarly, the initial values for smoking are generated based on the FinHealth 2017 study, and the distribution of smokers has changed significantly in Finland from 1992 when the MONICA data were collected. Furthermore, our definition of a smoker is not identical to the original FINRISK study \citep{VARTIAINEN2016213} where a person was considered a smoker if they had smoked for at least a year and had smoked in the last month. In the study, it is also not reported whether only cigarette smokers were considered, or if for example cigar and pipe smokers were included as well. The MONICA data do not provide enough information on smoking history of the participants to fully match the FINRISK definition.

In general, constructing a daily level simulator is challenging if only long-term prediction models are available. In a more ideal setting, we would have access to survival models, and thus more accurate information on the relation between the risk and the risk factors since no information would be lost via discretization into binary outcomes. Additionally, data containing in\textbf{}formation on the joint distribution of the risk factors is crucial in order to replicate the correlation structure accurately in the synthetic intitial population.

\subsection{Simulation results on the effect of interventions} 

We demonstrate the use of interventions in the framework by considering the causal effect of salt intake on the risk of stroke mediated by blood pressure. The typical salt intake in many countries is 9--12 g/day while the recommended level is 5--6 g/day \citep{who2003diet,he2013effect}. On the basis of a meta-analysis, a 100 mmol (6 g of salt per day) reduction in 24 hour urinary sodium was associated with a fall in systolic blood pressure of 5.8 mmHg \citep{he2013effect}. We consider three scenarios: The first scenario corresponds to the full population sample without interventions. In the second scenario, it is assumed that the food industry reduces the salt content in all products and consequently the daily sodium intake is 1 g lower. The systolic blood pressure will then be on average 0.97 mmHg lower. In the third scenario, an intervention is targeted to individuals with high blood pressure defined here as systolic blood pressure at or above 140 mmHg. These individuals are advised to stop adding salt in cooking and at table. It is assumed that 50\% of the target group will follow the advice. On the basis of a recent analysis \citep{karvanen2020dosearch}, this intervention is expected to reduce systolic blood pressure on average by 2.0 mmHg  among the compliers. In the framework, both interventions are implemented as \texttt{ManipulationEvent} instances and applied via the \texttt{intervene} method at the start of the simulation.

\begin{figure}[ht]
  \centering
  \includegraphics[width=0.50\textwidth]{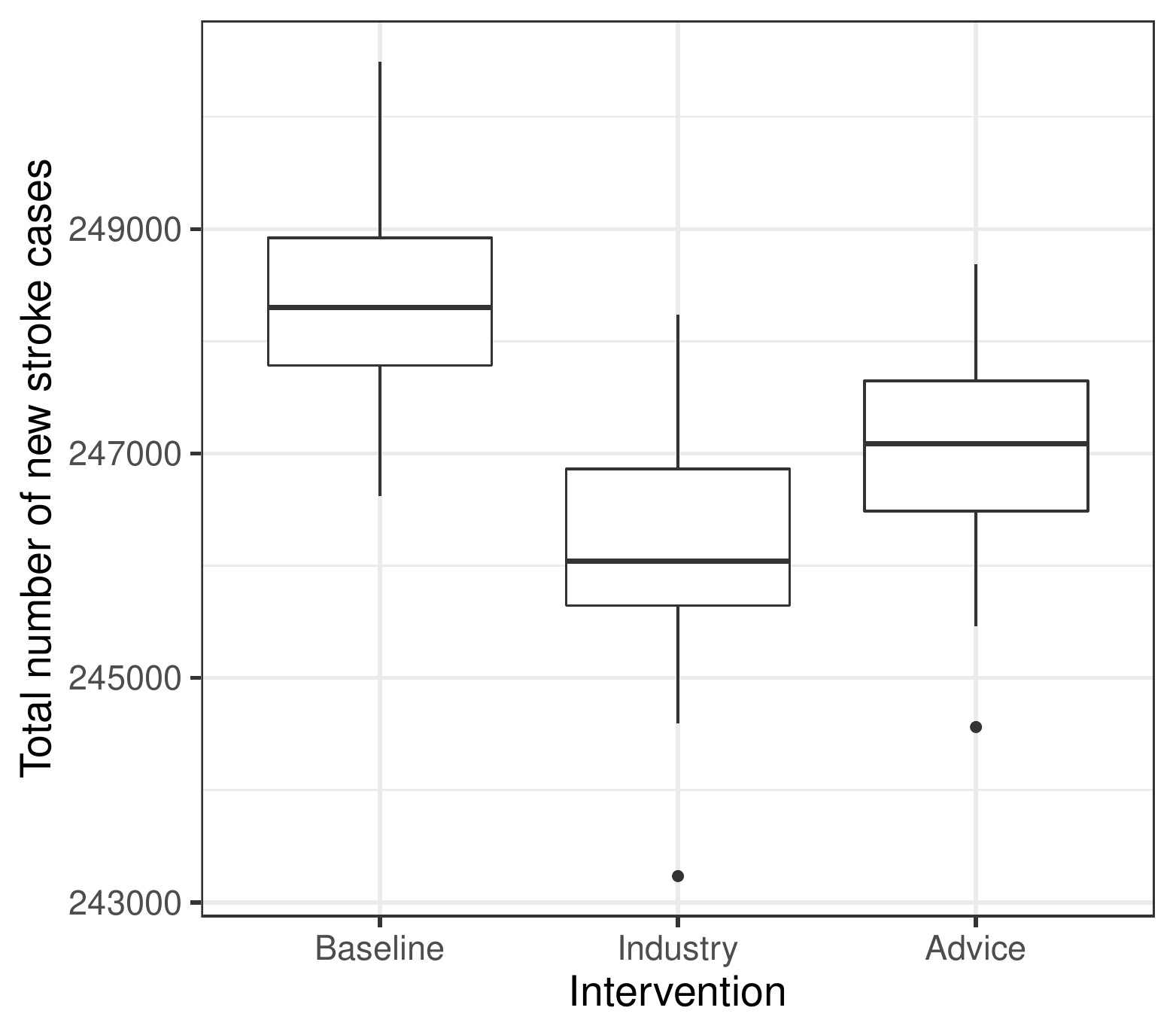}
  \caption{Boxplots of the total number of new stroke cases in the three scenarios from 100 replications of each scenario (from the same initial population in each scenario and replication). ``Baseline'' is the scenario without interventions, ``Industry'' corresponds to the salt content reduction carried out by the food industry and ``Advice'' is the recommendation regarding salt use targeted towards individuals with high blood pressure.}
  \label{fig:interventions}
\end{figure}

Figure~\ref{fig:interventions} shows the total number of new stroke cases in the three scenarios. We see that in this example the intervention where the food industry reduces the salt content in all products is the most effective at reducing the number of total stroke cases, while the advice to stop adding salt in cooking and at table targeted toward individuals with high blood pressure is slightly less effective in comparison.

\section{Discussion} \label{sec:discussion}

In this paper, we presented a framework for realistic data generation and simulation of complex systems and demonstrated its use in the heath domain. The framework implements functionalities that are in line with the features listed in Introduction. The framework supports individual level events on an arbitrary time scale and contains functions to calibrate the simulation parameters with external data sources. These functionalities together with sufficient knowledge for modeling data generating mechanisms are prerequisites for realistic simulations. The framework has modular structure which allows for simulating manipulations of different types.   The modular structure also makes it possible to simulate complex data collection processes. The state-of-the-art data structures, efficient random number generation and in-built parallelization ensure the scalability of the framework making it possible to run daily-level simulations for populations of millions individuals for decades.  The simulation runs are fully reproducible. An open-source implementation in R makes the framework directly available for statistics and data science communities.

The main use cases of the simulation framework are predicting the development of risk factors and disease occurrence,  evaluating the impact of interventions and policy decisions, and statistical method development. In the first two use cases, the support for calibration, the support for manipulations and the support for scalability are important features. The support for manipulations, especially the support for interventions, is crucial for medical decision making. In method development the role of data collection mechanisms becomes central. The simulation framework may also prove useful in other application areas outside of the health domain where decision making and the data collection functionality are of importance.



In future work, we would like to apply the presented framework to real-world scenarios similar to our example in Section~\ref{sec:examples} where decision making involving healthcare policies and treatment interventions is required. A key challenge for the use of simulations in medical decision making is related to causality. Information on the causal effects of risk factors is needed to simulate the effect of interventions realistically. By default, risk prediction models based on observational data are not causal models.  This holds also for the FINRISK and the FINDRISC risk score models defined in Section~\ref{sec:examplemodels}. The advances in causal inference and epidemiology are expected to improve the possibilities to conduct increasingly realistic simulations in the future.

\section*{Acknowledgments}
This research was supported by the Aca\-demy of Finland (grant no. 311877) and is related to the thematic research area DEMO (Decision Analytics Utilizing Causal Models and Multiobjective Optimization, \href{https://www.jyu.fi/it/en/research/research-projects/academy-of-finland/demo}{jyu.fi/demo}) of the University of Jyv\"askyl\"a, Finland. The authors wish to acknowledge CSC -- IT Center for Science, Finland for computational resources. 

\appendix

\section{Related Work} \label{app:literature}

There exist a large number of review papers considering the usage of simulation in the
healthcare domain. An umbrella review by \citet{Salleh2017} summarizes 37 surveys and divides the original works into four categories: \emph{types of applications}, \emph{simulation techniques used}, \emph{data sources}, and \emph{software used for simulation modelling}. 
According to this classification, our work falls under \emph{medical decision making} when considering the application domain. A considerable amount of literature has been published on simulation for medical decision making applications. Table~\ref{tab:relatedWorkMedicalDM} shows a selection of these studies, identified by going through the references of those reviews that according to the umbrella review by \citet{Salleh2017} reviewed the most medical decision making application studies \citep{katsaliaki2011applications,mielczarek2016review,mustafee2010profiling}. 
As summarized in the table, the models developed for this application area are typically not open-source. Moreover, in comparison to our work, most of these related studies dealt with only one particular decision making application for one particular country.  

At the same time, there is a relatively small body of literature that is concerned with open-source microsimulation frameworks within the medical domain (a summary of this literature is provided in Table~\ref{tab:relatedWorkOSframeworks}). 
In fact, \citet{spielauer2007dynamic} pointed out that microsimulation techniques largely focus on simulating tax-benefit and pension systems and that extensions for health-related policies and interventions are sparse. Although this situation has changed since 2007, the related work on open-source frameworks for this domain is still limited.
For example, a Scopus search in May 2020 with the TITLE-ABS-KEY search terms (open  AND source  AND framework  AND microsimulation  AND health) yielded only two results \citep{prakash2017cmost,kuchenbecker2018estimating}.
Similarly, a PubMed search with the same keywords also returned only two studies \citep{prakash2017cmost,sai2019multiobjective}. As documented in Table~\ref{tab:relatedWorkMedicalDM}, \citet{prakash2017cmost} developed and used a specific open-source tool for colorectal cancer microsimulation that was later used by \citet{sai2019multiobjective}, while \citet{kuchenbecker2018estimating} used an existing model to study pneumococcal diseases. In sum, these studies are limited to specific health problems and only Prakash et al. developed an open-source tool. The remaining studies in Table~\ref{tab:relatedWorkOSframeworks} were identified with a Google Scholar search with the same keywords specified above and by using the snowballing technique \citep{wohlin2014guidelines} among all already identified articles. 

The first three frameworks listed in Table~\ref{tab:relatedWorkOSframeworks} are all for general purposes (not specifically health related) open-source simulation frameworks that---like our framework---support large-scale individual-level data generation. However, in comparison to our framework, there is no support for completely synthetic data generation.  
The next framework in the table, \texttt{simSALUD}, is meant to simulate only small area health related issues. 
The last rows of the table describe the closest existing tools to our frameworks. 
The \texttt{microsimulation package in R} and \texttt{openM++} are available in R. However, they are not as scalable as our framework and do not include direct support for manipulations.

As summarized by \citet{krijkamp2018microsimulation}, R is appropriate not only for developing microsimulation models but also for the open-source distribution of the implementations. They emphasize that despite these advantages, open source models for medical decision making in R are scarce. 
To encourage more developers to use R, they provide an easy microsimulation example of a medical decision model (i.e., the sick-sicker model, first introduced by \cite{enns2015identifying}).
We compared the performance of \texttt{Sima} with their implementation. Running their vectorized example (by sourcing directly from GitHub and using the authors' own measure of running time) took $\approx 6$ seconds and running the \texttt{Sima} version took $\approx 2$ seconds if we record the entire simulation history (our implementation of the sick-sicker model is available at the Sima website \url{https://santikka.github.io/Sima/}).
Thus, we conclude that to our knowledge, no open-source simulation framework in R for realistic large-scale individual-level data generation for medical decision making was developed so far that is as scalable as \texttt{Sima}.

\begin{table*}[!ht]
    \centering
   \begin{scriptsize}
     \caption{Selection of related work on using simulation for medical decision making applications. Typically, models are not made open-source and developed only for one particular country and decision making problem.}
    \label{tab:relatedWorkMedicalDM}
    \begin{tabular}{p{0.1\linewidth}|p{0.36\linewidth}|p{0.36\linewidth}}
    	\hline
		Authors & Simulation Model, Objective   &  Software, Tools, Open-source Availability \\ 
		\hline
\citet{eldabi2000simulating}
& Discrete event simulation to simulate economic factors in adjuvant breast cancer treatment in England. 
& A package called \texttt{ABCSim} was created using the commercial simulation software \texttt{Simul8}. \texttt{ABCSim} is not open-source.
\\ 
\hline
\citet{cooper2002development}
& Discrete event simulation to model the progress of English patients who have had a coronary event through their treatment pathways and subsequent coronary events. 
& The simulation was written using the POST (Patient Oriented Simulation Technique) software with a Delphi interface. The developed model is not open-source.\\ 
\hline
\citet{caro2006budgetary} 
& Discrete event simulation to simulate the course of individuals with acute mania in bipolar I disorder and estimation of the budget impact of treatments for them from a United States healthcare payer perspective.
& The model is not open-source and the article does not provide any information on the used software.
\\ 
\hline
\citet{ahmad2007limiting}
& A 75-year dynamic simulation model comparing the long-term health benefits to society of various levels of tax increase to a viable alternative: limiting youth access to cigarettes by raising the legal purchase age to 21 in the United States.
& Vensim (i.e., commercial software) was used to develop a dynamic simulation model for estimating the population health outcomes resulting
from raising taxes on cigarettes and raising the legal smoking age to 21. The developed model is not open-source.
\\
\hline
\citet{huang2007cost} 
& A Monte Carlo simulation model to estimate the cost-effectiveness of improving diabetes care in federally qualified community health centers in the United States.
& Microsoft Excel 2000 and @Risk 4.5.4 for Windows were used to conduct the simulations and an older diabetes complications model \citep{eastman1997model} was adapted. The model is not open-source but the inputs are described in the article.
\\
\hline
\citet{zur2016microsimulation} 
& Microsimulation model of alcohol consumption and its effects on alcohol-related causes of death in the Canadian population. 
 Estimation of the cost-effectiveness of implementing universal alcohol screening and brief intervention in primary care in Canada. 
& The model was programmed and run in C and the results analyzed in R version 3.1.2. The programmed model is not open-source.
\\
\hline
\citet{larsson2018microsimulation}
& Individual-based microsimulation model to project economic consequences of resistance to antibacterial drugs for the Swedish health care sector.
& A dynamic microsimulation model developed by the Swedish Ministry of Finance 
called SESIM was used. Currently (May 2021), it is not possible to download SESIM. 
\\ 
\hline
\citet{prakash2017cmost,sai2019multiobjective} 
& \texttt{CMOST} is a microsimulation model for modeling the natural history of colorectal cancer, simulating the effects of colorectal cancer screening interventions, and calculating the resulting costs.  
& \texttt{CMOST} (Colon Modeling Open Simulation Tool) was implemented in Matlab and is freely available under the GNU General Public License at \url{https://gitlab.com/misselwb/CMOST}.  
\\
\hline 
\citet{kuchenbecker2018estimating}
& The authors adapted an existing microsimulation US model \citep{weycker2012public} 
to depict lifetime risks and costs of invasive pneumococcal diseases and nonbacteremic pneumonia, as well as the expected impact of different vaccination schemes, in a hypothetical population of German adults.
& No link to a documentation of the model or open-source framework can be found in the articles \citep{weycker2012public,kuchenbecker2018estimating}.
\\
\hline
\citet{spielauer2019portable} 
& \texttt{DYNAMIS-POP} is a dynamic micro-simulation model for population and education projections and for the simulation of policies. Adaptation  of  the  model  to  a  specific  country  only  requires  adapting  a  single  setup  script  and  simulation  module.  The  model is provided with test data of an imaginary country. 
 &  All components of \texttt{DYNAMIS-POP} including its code and all statistical analysis files are freely available and documented online at \url{http://dynamis.ihsn.org/}. Most statistical analysis scripts and scripts for post-processing and visualization of the results are implemented in R. 
\\
\hline 
\citet{hennessy2015population,manuel2014projections,manuel2016alzheimer,wolfson1994pohem} 
& \texttt{POHEM} (POpulation HEalth Model) is a longitudinal microsimulation model of health and disease. The model simulates representative populations and allows the rational comparison of competing health intervention alternatives, in a framework that captures the effects of disease interactions. 
& \texttt{POHEM} was developed in Modgen (see Table~\ref{tab:relatedWorkOSframeworks}). However, the \texttt{POHEM} model implementation itself is not open-source (and so far still limited to Canada, i.e., one country).  
\url{https://www.statcan.gc.ca/eng/microsimulation/modgen/new/mods/pohem}.
\\
\hline
\cite{caron2015demosim,marois2020implementing} 
& \texttt{DemoSim} is a microsimulation model developed and maintained at Statistics Canada that is designed to produce population projections. 
& The implementation is not open-source but documented at \url{https://www.statcan.gc.ca/eng/microsimulation/demosim/demosim}.
Variants of \texttt{DemoSim} were recently developed also for several European countries and Australia \citep{marois2020implementing}.
\\
\hline
\end{tabular} 
\end{scriptsize}
\end{table*}

\begin{table*}[!ht]
\centering
\begin{scriptsize}
\caption{Related work on open-source frameworks for microsimulation models.}
\label{tab:relatedWorkOSframeworks}
\begin{tabular}{p{0.1\linewidth} | p{0.44\linewidth} |p{0.28\linewidth}}
\hline
Authors & Simulation Model/ Objective   &   Link / Open-source Availability \\ 
\hline
\citet{deMenten2014liam2} 
& \texttt{LIAM2} is an open-source development tool for discrete-time dynamic microsimulation models. The framework is for general purposes (not specifically health related). It allows for the simulation of discrete-time dynamic models with a cross-sectional time step of whatever kind of objects the modeler chooses. It provides functionalities to calibrate to exogenous information of any number of dimensions. 
& \texttt{LIAM2} was developed primarily in Python, it can be downloaded from \url{http://liam2.plan.be}. It is licensed under the GNU General Public License, meaning one can freely use, copy, modify and redistribute the software.
\\
\hline 
\citet{mannion2012jamsim} 
& \texttt{JAMSIM} (JAva MicroSIMulation) is a synthesis of open-source packages that provides an environment and set of features for the creation of dynamic discrete-time microsimulation models that are to be executed, manipulated and interrogated by non-technical, policy-oriented users. 
& \texttt{JAMSIM} is freely available as an open-source tool, for public reuse and modification at \url{http://code.google.com/p/jamsim/}. It combines R and the Java-based agent-based modeling graphical tool Ascape.
\\
\hline 
\citet{richiardi2017jas}
& \texttt{JAS-mine} is a Java-based computational platform that features tools for discrete-event simulations encompassing both dynamic microsimulation and agent-based modeling. Objectrelational mapping is used to embed a relational database management system. It is a general-purpose platform (not specifically health related).
& \texttt{JAS-mine} is freely available as an open-source tool, for public reuse and modification at \url{https://github.com/jasmineRepo/JAS-mine-core}. It is written in Java.
\\
\hline 
\citet{kosar2014simsalud, tomintz2017simsalud}
& \texttt{simSALUD} is a deterministic spatial microsimulation framework. Its main aim is to model health related issues for small areas using spatial microsimulation modeling to simulate information where no data exists or is accessible.
& \texttt{simSALUD} is an open-source web-application, programmed using Java and based on Apache’s model-view-controller application Struts2. It can be accessed at \url{http://www.simsalud.org/simulation/}.
\\ 
\hline 
\citet{clements2021discrete}
& \texttt{microsimulation} package for R provides implementations of discrete event simulations in both R and C++. The R implementation builds on several R5 classes (useful from a pedagogical perspective, but slow for larger microsimulations). For speed, the authors provide C++ classes.
& The package is freely available at \url{https://github.com/mclements/microsimulation}. 
\\
\hline	
Developed by Statistics Canada; used, for example, in \citet{spielauer2019portable}
& \texttt{Modgen} and its open-source implementation \texttt{openM++} are generic microsimulation software supporting the creation, maintenance and documentation of dynamic microsimulation models. Several types of models can be accommodated, be they continuous or discrete time, with interacting or non-interacting populations. Compared to its closed source predecessor \texttt{Modgen}, \texttt{openM++} has many advantages like portability, scalability and open source. 
& \texttt{openM++} is freely available at \url{https://github.com/openmpp}.
\\
\hline	
\end{tabular} 
\end{scriptsize}
\end{table*}

\section{Mathematical Details} \label{app:maths}

We present here a mathematical formulation of our simulation framework. The implementation of Sima builds on these concepts and definitions.

Simulation domain defines the scope of the status variables, parameters and latent variables.
\begin{definition}[Simulation domain] A \emph{simulation domain} is a triple
\[
  (\mathcal{X}, \Theta, \Psi),
\]
where $\mathcal{X} = \bigtimes_{j=1}^m \mathcal{X}_j$ is the space of the status variables, $\Theta$ is the parameter space and $\Psi$ is the latent variable space.
\end{definition}
We assume that the spaces $\mathcal{X}_j$ of individual status variables are one dimensional and real-valued. This assumption does not rule out for example categorical variables, which can be included by encoding them in the standard way with a binary indicator for each category. It should be noted we do not impose any restrictions on the possible combinations of values in the Cartesian product $\mathcal{X}$.

\begin{definition}[Population] Let $\mathcal{D} = (\mathcal{X}, \Theta, \Psi)$ be a simulation domain. A \emph{population} $\+ X$ in $\mathcal{D}$ is a $n \times m$ matrix
\[
  \+ X = \begin{bmatrix}
    \bs x_1^\tpose \\
    \vdots \\
    \bs x_n^\tpose
  \end{bmatrix} = \begin{bmatrix}
  x_{11} & \cdots & x_{1m} \\
  \vdots & \ddots & \vdots \\
  x_{n1} & \cdots & x_{nm}
  \end{bmatrix},
\]
where each row vector $\bs x_i^\tpose$ refers to the \emph{individual} $i$ of the population, $n$ is the population size, and $x_{ij} \in \mathcal{X}_j$ for all $i = 1,\ldots,n,\, j = 1,\ldots,m$. The element $x_{ij}$ is called the \emph{status variable} $j$ of individual $i$. 
\end{definition}
The set of populations of size $n$ having $m$ status variables in $\mathcal{D}$ is denoted by $\mathcal{P}^{n \times m}[\mathcal{D}]$. Changes in the population are carried by the event functions.


\begin{definition}[Manipulation event] Let $\mathcal{D} = (\mathcal{X}, \Theta, \Psi)$ be a simulation domain. A \emph{manipulation event} $\mu$ in $\mathcal{D}$ is a function $\mu:\mathcal{X} \times \Theta \times \Psi \rightarrow \mathcal{X}$.
\end{definition}
Manipulation events are used to describe deterministic changes between two time points in the status variables of a single individual $\bs x_i^\tpose$ of a population $\+ X$. We also define the following. 

\begin{definition}[Accumulation event] Let $\mathcal{D} = (\mathcal{X}, \Theta, \Psi)$ be a simulation domain. An \emph{accumulation event} in $\mathcal{D}$ is a pair
\[ (\eta, \nu), \]
where $\eta$ is a function $\eta:\Theta \times \Psi \rightarrow \mathbb{N}$ and $\nu$ is a sequence $\{b_k\}_{k=1}^\infty$ of functions $b_k:\Theta \times \Psi \rightarrow \mathcal{P}^{k\times m}[\mathcal{D}]$. 
\end{definition}

The operation of accumulation events is twofold. First, the function $\eta$ selects the number of individuals $k$ to be added to the population. Second, the corresponding function $b_k$ is used to generate the new set of individuals to be added. The case where $\eta(\bs \theta, \bs \psi) = 0$ for some assignment of $\bs \theta \in \Theta$ and $\bs \psi \in \Psi$ should be understood to mean that no new individuals will be added to the population. The events are defined as deterministic functions given the parameters $\bs \theta$ and the latent variables $\bs \psi$ which, if assumed to be random variables, give a stochastic nature to the simulator. Event order is defined as follows.



\begin{definition}[Event order] Let $\mathcal{D} = (\mathcal{X}, \Theta, \Psi)$ be a simulation domain and let $k \in \mathbb{N}$. An \emph{event order} in $\mathcal{D}$ of $k$ manipulation events is a set of functions $\{\sigma_1, \ldots, \sigma_n \}$, such that
\[
    \sigma_n : \mathbb{N}_k \times \Psi \rightarrow \mathbb{N}_k,\quad \textrm{ for all } n \in \mathbb{N},
\]
where $\mathbb{N}_k = \{1,\ldots,k\}$.
\end{definition}

A single function $\sigma_j$ is essentially a permutation function that defines the application order of manipulation events for individual $j$, which can be random, as these functions depend on the latent variables $\Psi$. We are now ready to define the simulator itself.

\begin{definition}[Simulator] A \emph{simulator} is a quintuple
\[ (\mathcal{D}, \+ X_0, \{\mu_i\}_{i = 1}^r, E_r, A), \]
where $\mathcal{D}$ is a simulation domain, $\+ X_0$ is the initial population in $\mathcal{D}$, $\{\mu_i\}_{i = 1}^r$ is a sequence of manipulation events in $\mathcal{D}$, $E_r$ is an event order in $\mathcal{D}$ of $r$ manipulation events and $A$ is a set of accumulation events in $\mathcal{D}$.
\end{definition}

At a given time, the simulator occupies a state.


\begin{definition}[State] Let 
\[
    \mathcal{S} = ((\mathcal{X}, \Theta, \Psi), \+ X_0, \{\mu_i\}_{i = 1}^r, E_r, A)
\]
be a simulator. A \emph{state} of $\mathcal{S}$ is a pair
\[
    (\+ X, \bs \theta),
\]
where $\+ X$ is a population in $(\mathcal{X}, \Theta, \Psi)$ and $\bs \theta \in \Theta$. A state of $\mathcal{S}$ is its initial state if $\+ X = \+ X_0$. The set of all states of $\mathcal{S}$ is denoted $\Sigma[\mathcal{S}]$.
\end{definition}

A state of the simulator encapsulates the status of its population ($\+ X$) and its parameter configuration ($\bs \theta$). Starting from an initial state, transitions between states are carried out by the transition function.

\begin{definition}[Transition function] Let 
\[\mathcal{S} = ((\mathcal{X}, \Theta, \Psi), \+ X_0, \{\mu_i\}_{i = 1}^r, E_r, A) \] 
be a simulator. The \emph{transition function} $\delta$ of $\mathcal{S}$ is a function $\delta : \Sigma[\mathcal{S}] \times \Psi \rightarrow \Sigma[\mathcal{S}]$ such that
\[
    \delta(S, \bs \psi) = \delta((\+ X, \bs \theta), \bs \psi)
    = (d(\+ X, \bs \theta, \bs \psi), \bs \theta)
    = (\+ X^\prime, \bs \theta),
\]
where $S \in \Sigma[\mathcal{S}]$ and $d(\+ X, \bs \theta, \bs \psi) = d_m(d_a(\+ X, \bs \theta, \bs \psi), \bs \theta, \bs \psi)$ is defined as follows
\[
    d_m(\+ X, \bs \theta, \bs \psi) = \bigcirc_{i = 1}^{r} u_i(\+ X, \bs \theta, \bs \psi), \quad
    u_i(\+ X, \bs \theta, \bs \psi) = \begin{bmatrix}
      \mu_{\sigma_1(i,\bs \psi)}(\+ x_1^\tpose, \bs \theta, \bs \psi) \\
      \vdots \\
      \mu_{\sigma_n(i,\bs \psi)}(\+ x_n^\tpose, \bs \theta, \bs \psi)
    \end{bmatrix}, \]
\[
    d_a(\+ X, \bs \theta, \bs \psi) = \begin{bmatrix}
      \+ X \\
      b_{\ell_1, k(\ell_1)}(\bs \theta, \bs \psi) \\
      \vdots \\
      b_{\ell_s, k(\ell_s)}(\bs \theta, \bs \psi)
      \end{bmatrix},
\]
where $(\alpha_j, \{b_{j,k}\}_{k=1}^\infty) \in A$, $k(j) = \alpha_{j}(\bs \theta, \bs \psi)$, $\{\ell_1,\ldots,\ell_s\} = \{j \,|\, j = 1,\ldots,|A|,\, k(j) > 0\}$ and $\sigma_j \in E_r$. The symbol $\bigcirc_{i=1}^m$ denotes an ordered composition of a sequence of functions $\{f_i\}_{i=1}^m$ over the leftmost argument, i.e., $\bigcirc_{i=1}^m f_i = f_1 \circ f_2 \circ \cdots \circ f_m$.
\end{definition}

Given a latent vector $\bs \psi$, the transition function defines the transformation from one state to the next. The functions $d_m$ and $d_a$ correspond to the transformations carried out by manipulation events and the accumulation events. Since accumulation events do not depend on the current population, the order in which they are applied does not matter in $d_a$. However, the order is significant for manipulation events in $d_m$. Furthermore, this order can vary between individuals if the elements of the event order are defined as $\sigma_j(i,\bs \psi)$ instead of just $\sigma(i,\bs \psi)$, which would indicate a shared order for all individuals instead.

A single run of the simulator provides the user with a simulation sequence.

\begin{definition}[Simulation sequence]\label{def:sim_seq} Let \[\mathcal{S} = ((\mathcal{X}, \Theta, \Psi), \+ X_0, \{\mu_i\}_{i = 1}^r, E_r, A) \] be a simulator, $S_0 \in \Sigma[\mathcal{S}]$, and let $\{\bs \psi_t\}_{t=0}^T$ be a sequence of $(\Psi, \mathcal{F})$-measurable random vectors. A simulation sequence of $\mathcal{S}$ is a sequence of its states $\{S_t\}_{t=0}^T$ such that
\[
S_t = \begin{dcases*}
  S_0 & if $t = 0$, \\
  \delta(S_{t-1},\bs \psi_t) & if $t > 0$,
\end{dcases*}
\]
where $\delta$ is the transition function of $\mathcal{S}$.
\end{definition}
The index $T$ defines the length of the simulation sequence. In practice, $T$ can be viewed as a length of a specific time interval and incrementation of time $t$ by 1 can be understood to mean the progress of time at the desired granularity (in days, weeks, years, etc.). Note that in a single simulation sequence the parameters $\bs \theta$ are determined by the state at time $t = 0$ and remain unchanged throughout the sequence.

In order to calibrate the simulator, we need to be able to change its parameter configuration while keeping the system otherwise intact. This feature is provided by the configuration function.

\begin{definition}[Configuration function] Let
\[\mathcal{S} = ((\mathcal{X}, \Theta, \Psi), \+ X_0, \{\mu_i\}_{i = 1}^r, E_r, A) \] be a simulator. The \emph{configuration function} $c$ of $\mathcal{S}$ is a function $c : \Sigma[\mathcal{S}] \times \Theta \rightarrow \Sigma[\mathcal{S}]$ such that
\[
    c(S, \bs \theta^\prime) = c((\+ X, \bs \theta), \bs \theta^\prime)
    = (\+ X, \bs \theta^\prime),
\]
where $S \in \Sigma[\mathcal{S}]$.
\end{definition}
As the name suggests, the configuration function $c$ can be used to configure the simulator, meaning that it can be used to manipulate the current state by exchanging the parameter vector $\bs \theta$ for $\bs \theta^\prime$. The data collection process of the simulator is provided by the sampler function.

\begin{definition}[Sampler] Let $\+ X$ be a population in $\mathcal{D}$. A \emph{sampler} is a function $s : \Psi \times \mathcal{P}^{n \times m}[\mathcal{D}] \rightarrow \mathcal{P}^{n \times m}[\mathcal{D}^*]$ such that
\[
  s(\bs \psi, \+ X) = \+ X^*,
\]
where $\+ X^*$ is called a \emph{sample} with either $x^*_{ij} = e(\bs \psi, x_{ij})_j$ or $x^*_{ij} = \textrm{NA}$ and $e_{j} : \Psi \times \mathcal{X}_j \rightarrow \mathcal{X}_j$ is an error function for all $i = 1,\ldots,n$ and $j = 1,\ldots,m$. NA denotes a missing value and $\mathcal{D}^*$ denotes a simulation domain extended to include missing values.
\end{definition}

\section{Efficient Pseudo-random Number Generation} \label{app:dqngr}

The built-in pseudo-random number generators of R are fairly efficient, but they can be improved upon, and significant reductions in simulation time can be observed when the default generators are replaced by more advanced methods. The importance of fast and scalable PRNGs is further highlighted for large populations and long simulation runs. The R package \texttt{dqrng} \citep{dqrng} provides several fast PRNG generators, including C\texttt{++} implementations of the permuted congruential generator (PCG) family by \citet{oneill:pcg2014}, and Xoroshiro128\texttt{+} and Xoshiro256\texttt{+} by \citet{blackman2019}. The Ziggurat algorithm proposed by \citet{JSSv005i08} is provided for generating random numbers from normal and exponential distributions. A version of the Threefry engine by \citet{10.1145/2063384.2063405} is also available. While cryptographic aspects of PRNGs may be of importance in some practical applications of the simulator, our focus is mainly in the performance.

Generators of the \texttt{dqrng} package are used internally in the implementation of the simulator, one example being the generation of the order in which manipulation events are applied at each time point. Typically, performance is improved over a base R generator even when using uniform random variables and the inverse method to generate realizations from a distribution for which no prespecified generator is available in the \texttt{dqrng} package. When manipulation events and accumulation events are defined for a specific simulator instance, the choice of PRNG for the random functionality of the events themselves can take advantage of the \texttt{dqrng} package but are not restricted to it. However, the user has to take care in order to ensure that the simulation remains reproducible if multiple different generators are used simultaneously.

\begin{figure*}[ht]
  \centering
  \includegraphics[width=0.85\textwidth]{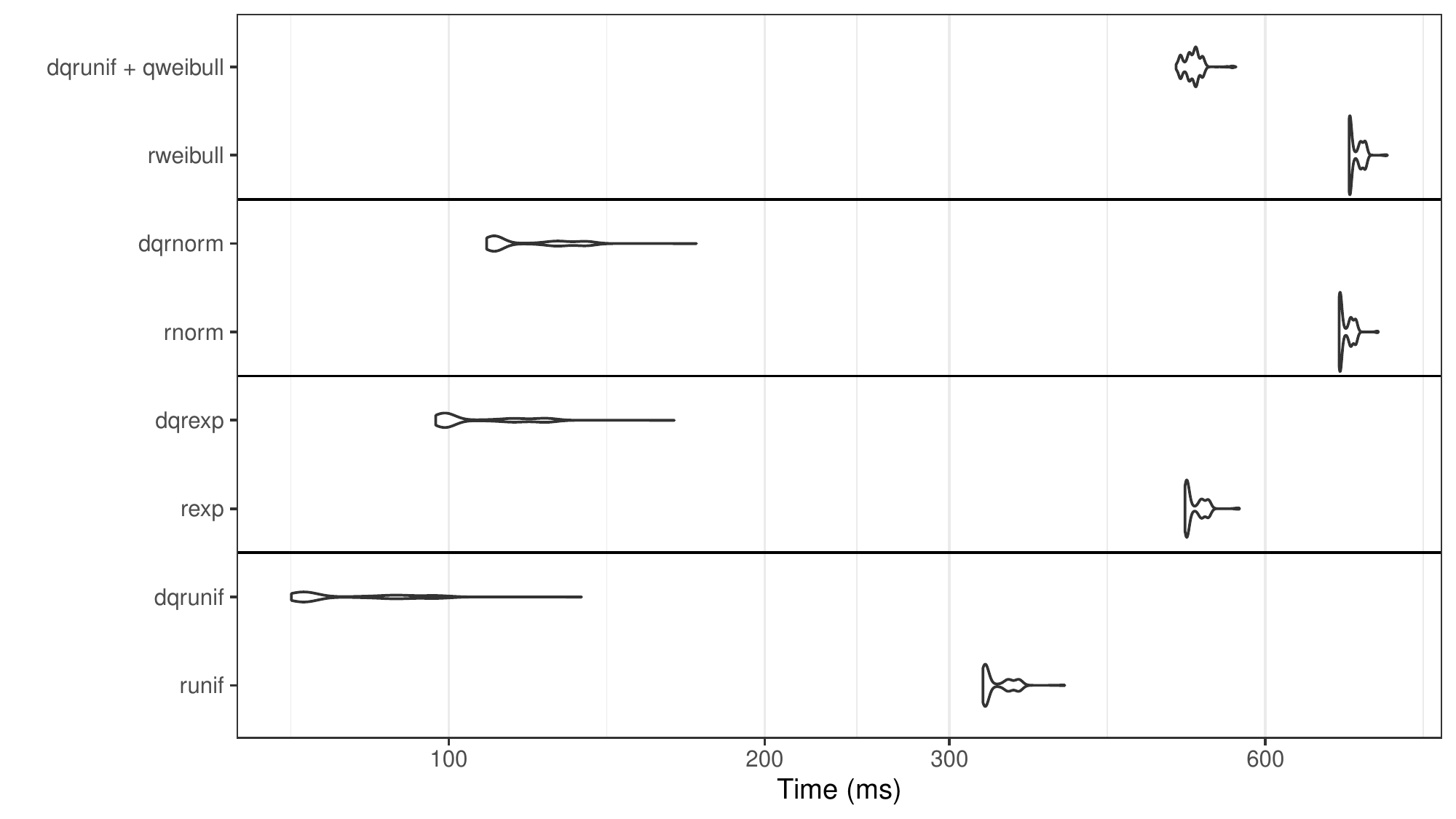}
  \caption{Density plots of time taken (in milliseconds) for 10 million random draws over 1000 replications for the Xoroshiro128\texttt{+} PRNG implementation of \texttt{dqrng} and the 32 bit Mersenne-Twister of base R with Weibull, normal, exponential and uniform distributions. Prefix ``dq'' denotes the \texttt{dqrng} version of the function. The horizontal axis uses $\log_{10}$-scaling. Weibull-distributed random variables are generated using the inverse method.}
\label{fig:dqrng}
\end{figure*}

We conduct a benchmark comparing Xoroshiro128\texttt{+} provided by \texttt{dqrng} to the base R 32 bit Mersenne-Twister \citep{10.1145/272991.272995}. The distributions included in \texttt{dqrng} consist of exponential, gaussian and uniform distributions. In addition, we compare the generation of Weibull-distributed random variables using the inverse method and uniform random variables with \texttt{dqrng} against the base R \texttt{rweibull} function. The results are presented in Figure~\ref{fig:dqrng} and they show that \texttt{dqrng} dominates the default generators for the included distributions of the package. The difference is smaller when using the inverse method with the quantile function \texttt{qweibull}, but \texttt{dqrng} still always outperforms the base generators.

\clearpage
\bibliographystyle{abbrvnat}
\bibliography{references,litReview}  

\end{document}